\documentclass[pdflatex,sn-mathphys-num]{sn-jnl}

\usepackage{graphicx}%
\usepackage{multirow}%
\usepackage{amsmath,amssymb,amsfonts}%
\usepackage{amsthm}%
\usepackage{mathrsfs}%
\usepackage[title]{appendix}%
\usepackage{xcolor}%
\usepackage{textcomp}%
\usepackage{manyfoot}%
\usepackage{booktabs}%
\usepackage{algorithm}%
\usepackage{algorithmicx}%
\usepackage{algpseudocode}%
\usepackage{listings}%

\theoremstyle{thmstyleone}%
\theoremstyle{thmstyletwo}%

\theoremstyle{thmstylethree}%

\begin{document}

\title[Phenomenological constraints on \newline"impossible measurements"]{Phenomenological constraints on "impossible measurements"}


\author*[1]{\fnm{Jesse} \sur{Huhtala}}\email{jejohuh@utu.fi}

\author[1]{\fnm{Iiro} \sur{Vilja}}\email{vilja@utu.fi}

\affil*[1]{\orgdiv{Department of Physics and Astronomy}, \orgname{University of Turku}, \orgaddress{ \city{Turku}, \postcode{20014},  \country{Finland}}}


\abstract{In this article, we analyze an "impossible measurement" scenario presented by Sorkin. This scenario involving a joint measurement on spacelike separated systems in an intermediary region has widely been discussed in the quantum field theory measurement literature. We analyze the non-relativistic version of this paradoxical measurement scenario in full detail and give explicit bounds for the amount of signaling present. We also discuss the conditions under which no extraneous signaling occurs.}




\maketitle

\section{Introduction}\label{sec:intro}
It is well-known that a naive, straightforward application of the projection postulate for measurement leads to faster-than-light signaling, as was shown by Sorkin in \cite{sorkinImpossibleMeasurementsQuantum1993}. Sorkin's impossible measurements have been often cited in relativistic measurement theory; this problem has been analyzed by many authors in the QFT case \cite{fewsterLecturesMeasurementQuantum2025,ramonRelativisticCausalityParticle2021,papageorgiouEliminatingImpossibleRecent2024,borstenImpossibleMeasurementsRevisited2021,bostelmannImpossibleMeasurementsRequire2021,polo-gomezDetectorbasedMeasurementTheory2022} both in terms of detector models (e.g. \cite{polo-gomezDetectorbasedMeasurementTheory2022, deramonCausalitySignallingNoncompact2023}) and in the algebraic framework (e.g. \cite{hellwigFormalDescriptionMeasurements1970, fewsterQuantumFieldsLocal2020}).

Our aim is to investigate Sorkin's "impossible measurements"\ in the non-relativistic case to determine the impact of Sorkin's scenario on signaling. Most of the previous work on this has focused on quantum field theory (though see \cite{gisinMeasurementTheoryQFT2024, blochRelativisticOdditiesQuantum1967}). There are well-established no-go theorems for faster-than-light signaling \cite{eberhardQuantumFieldTheory1989,ghirardiGeneralArgumentSuperluminal1980}, so our question is: what is it that faster-than-light signaling measurement scenarios like Sorkin's actually add to the signaling already present in the non-relativistic theory \cite{sorkinImpossibleMeasurementsQuantum1993}? 

In non-relativistic quantum mechanics, a wave function that is localized at time $t=0$ will always have non-zero support everywhere at any later time after undergoing Schrödinger evolution; at a later time, for a single free particle of mass $m$ in d-dimensional space,
\begin{align}
    \psi(x,t) = \Theta (t-t_0) \bigg( \frac{m}{2\pi i(t-t_0)} \bigg)^{d/2}\int_{\mathbb{R}^d} d^dx' \exp \bigg( \frac{im|x-x'|^2}{2(t-t_0)} \bigg)\psi _0(x,t_0).
\end{align}
Thus, one might question the wisdom of examining the signaling properties of a non-relativistic theory in the first place. The question is interesting because of the popularity of Sorkin's scenario and the continuing interest in it \cite{polo-gomezStateUpdatesUseful2025}. While non-relativistic theories tend to always signal in any case, there are still limits (of e.g. the Lieb-Robinson type \cite{liebFiniteGroupVelocity1972a}) to the "amount"\ of signaling present; hence it is reasonable to ask, even in the context of a non-relativistic theory, whether Sorkin's scenario enhances this signaling or not. There are also various relativistic theorems that tend to be applied even in otherwise non-relativistic calculations, such as the no-signaling theorem \cite{ghirardiGeneralArgumentSuperluminal1980}; hence it is reasonable to ask whether Sorkin's scheme makes these theorems less applicable to other similar situations. The scenario also elegantly captures many of the elements in a theory that make signaling possible. It is therefore worthy of analysis.

Let us now recall Sorkin's original argument \cite{sorkinImpossibleMeasurementsQuantum1993}. Sorkin considers three spacetime regions $O_1$, $O_2$ and $O_3$ such that some points of $O_1$ causally precede $O_2$ and some points of $O_2$ precede $O_3$. We follow \cite{sorkinImpossibleMeasurementsQuantum1993} in taking $O_1$ to be in $O_2$'s causal past and $O_2$ in $O_3$'s past without loss of generality. Regions $O_1$ and $O_3$ are taken to be spacelike separated. The system under study is a pair of spins, initially both in the down-state, which we call $|dd\rangle$. In $O_1$, a kick is performed on the first spin, transforming the first spin in to the up-state, thus the total state is $|ud\rangle$. An observer in $O_2$ performs a joint projective measurement along
\begin{align}
    \hat{B} = \frac{1}{\sqrt{2}}(|uu\rangle + |dd\rangle)
\end{align}
and finally, a person in the region 3 measures an observable $\hat{C}$ on the second spin.

Sorkin shows that, in case no kick happened, the expectation value of $\hat{C}$ is 
\begin{align}
    \langle \hat{C}\rangle = \frac{1}{2}\text{tr}\hat{C}.
\end{align}
On the other hand, if there was a kick, the expectation value is
\begin{align}
    \langle\hat{C}\rangle = \langle d|\hat{C}|d\rangle,
\end{align}
where the inner product is in the Hilbert space of the second spin. This clearly provides faster-than-light signaling between regions 1 and 3, which we presumed to be spacelike separated since the presence of the second spin in state $|u\rangle$ is an unambiguous signal that no kick happened. As Sorkin points out, assuming a "kick"\ operation is not necessary; a similar result obtains if we perform a measurement in region 1.

Sorkin used this non-relativistic version merely as a prelude to the QFT case, and as such, he neglected to model certain important parts of the scenario. We will follow Sorkin's basic idea: the presence of a spin-up in the last region, region $O_3$, will be a sign of signaling. In order to investigate the signaling properties of the theory, we will have to add spatial variables to our calculation; one should not speak of spacetime regions without modeling them in the quantum-mechanical description itself. We will take the two spin particles to be fermionic systems, which are antisymmetric under exchange of particles, and seek unitary measurement operations that preserve this antisymmetry. We could also choose a bosonic state, but this makes no real difference to the results.

Our strategy is straightforward. We include the spatial variables and the antisymmetry in to the calculation that Sorkin performed. That will, of course, mean that there is no way to "kick the second particle"\ or to "measure the first particle", since the particles are indistinguishable. Further, since the theory is non-relativistic, there is no way for two regions to be fully causally separated. Consequently, there will be some non-zero probability to obtain a spin-up result in the $O_3$ measurement even with the kick. We will instead place the regions $O_1$ and $O_3$ sufficiently far apart that the probability to obtain a spin-up measurement without the intermediate measurement in $O_2$ is negligible (corresponding to the kicked case), as is usually implicitly done when using the no-signaling theorem in a non-relativistic context. We will, in fact, derive an upper bound for this probability in terms of Heisenberg-picture projection operators. Any probability of spin-up smaller than this upper bound will be considered to not contain useful signaling.

It should be clear that these results will heavily depend on how the measurement in $O_2$ is implemented. We will first derive a bound for a relatively general scenario using spatial projections for the measurement apparatuses, and then discuss other possible realizations. We will then specialize to a particular time evolution to get some concrete results.

\section{Analysis of the scenario}\label{sec:analysis}
We will work in the Hilber space
\begin{align}
    \mathcal{H} = \mathcal{H}_{\psi}\otimes \mathcal{H}_{O_2} \otimes \mathcal{H}_{O_3} 
\end{align}
where $\mathcal{H}_{\psi}$ is the antisymmetric subspace of $l^2\otimes l^2 \otimes \mathbb{C}^2\otimes \mathbb{C}^2$ (i.e. the spatial and spin parts of the wave function), $H_{O_2} = \mathbb{C}^2$ is the $O_2$ detector modeled by a single qubit, and $H_{O_3} =\mathbb{C}^2$ is the $O_3$ detector also modeled by a single qubit. The sequence of operations is such that at time $t=0$, we have the particle in region $R_1$, where it is either kicked or not. Some time $t_1$ passes, during which the wave function evolves as a result of unitary $\hat{U}(t_1)$,  and the measurement in $O_2$ is then performed. Afterwards, the wave function evolves by $\hat{U}(t_2)$, and finally the $O_3$ measurement is performed at time $t_3 := t_1+t_2$.

Let us define some useful notation. We take the spacetime region $O_2$ to consist of a time $t_2$ and a space region $R_2$, and similarly for $O_3$ (i.e. the measurement time is sufficiently small to be negligible). We then write, in the Heisenberg picture,
\begin{align}
    P^{O_i} = P^{R_i}(t_i).
\end{align}
In addition, the lower index on a projection operator will refer to which subspace of the spatial Hilbert space is being referred to, i.e. on the spatial subspace $l^2\otimes l^2$ where $l^2$ is the Lebesgue space. On that space, $P_1 = P \otimes 1$, $P_2 = 1\otimes P$.
\begin{align}
    P^R_{00} &:= (1-P^{R}_1)(1-P_2^{R}),\\
    P^R_{01} &:= (1-P^{R}_1)P_2^{R},\\
    P^R_{10} &:= P_1^{R}(1-P_2^{R}),\\
    P^R_{11} &:= P_1^{R}P_2^{R}.
\end{align}

With these notations, we can define our measurement operators. First, the measurement operator on $O_3$ is implemented by the unitary (on the spatial, spin and $O_3$ qubit subspace)
\begin{align}
    \hat{U}^{O_3} := &P^{O_3}_{00}\otimes1_{O_2}\otimes1_{\text{spins}}\otimes1_{O_3} + P^{O_3}_{10}\otimes 1_{O_2}\otimes \hat{M}^{(1)} \nonumber\\
    &+ P^{O_3}_{01}\otimes 1_{O_2}\otimes \hat{M}^{(2)}+P^{O_3}_{11}\otimes1_{O_2}\otimes1_{\text{spins}}\otimes1_{O_3} 
\end{align}
with
\begin{align}
    \hat{M}^{(1)}:= \begin{cases}
    |ds0\rangle &\mapsto |ds0\rangle\\
    |us1\rangle &\mapsto |us1\rangle \\
    |ds1\rangle &\mapsto |us0\rangle \\
    |us0\rangle &\mapsto |ds1\rangle
    \end{cases},\quad \hat{M}^{(2)}:= \begin{cases} |sd0\rangle &\mapsto |sd0\rangle\\
    |su1\rangle &\mapsto |su1\rangle \\
    |sd1\rangle &\mapsto |su0\rangle \\
    |su0\rangle &\mapsto |sd1\rangle
    \end{cases},
\end{align}
where $s$ is an arbitrary spin state. In other words, the operator $M^{(i)}$ transfers one of the two spin variables in to the measurement qubit in $O_3$; the index $i$ refers to which spin state is being transferred. The $O_2$ measurement is
\begin{align}
    \hat{U}^{O_2} := (1-P^{O_2}_{11}) \otimes 1_{\text{spins}} \otimes 1_{O_2}\otimes 1_{O_3} + P^{O_2}_{11}\otimes \bigg[ (1_{\text{spins}}-\Pi^+)\otimes 1_{O_2} \otimes 1_{O_3} + \Pi ^+ \otimes X \otimes 1_{O_3} \bigg]
\end{align}
with
\begin{align}
    \Pi^+ = \frac{1}{2}(|uu\rangle + |dd\rangle)(\langle uu| + \langle dd|) 
\end{align}
the projective spin measurement in $O_2$. With these definitions, a "spin-up"\ measurement in the region $O_3$ means that the qubit, initially in state $|0\rangle$, has flipped to $|1\rangle$.
\subsection*{The kicked scenario}\label{sec:kick}
The scenario in which the particle is in region $R_1$ initially and is furthermore kicked provides our baseline. The result of this scenario will be used to derive an upper bound for the probability of obtaining a spin-up result; any probability below this bound is considered negligible. We start from the state
\begin{align}
    |\psi_{\text{kick}}\rangle = [|\psi\rangle \otimes|ud\rangle 
    - |\tilde{\psi}\rangle\otimes|du\rangle]\otimes |0\rangle_{O_2}\otimes |0\rangle_{O_3},
\end{align}
where we define
\begin{align}
    |\psi\rangle := P_1^{O_1}\psi(x_1,x_2), \qquad
    |\tilde{\psi}\rangle := P_2^{O_1}\psi(x_2,x_1).
\end{align}
Applying the described sequence of operations (the details are in appendix \ref{app:tech}), we arrive at the probability $p_1^{\text{kick}}$ of finding the final measurement qubit in state $|1\rangle$:
\begin{align}
    p_1^{\text{kick}} = || P^{O_3}_{10} |\psi\rangle||^2 +||P^{O_3}_{01}|\tilde{\psi}\rangle||^2 =  ||P^{O_3}_{10}P^{O_1}_1\psi(x_1,x_2)||^2 + ||P^{O_3}_{01}P^{O_1}_2\psi(x_2,x_1)||^2
\end{align}
from which we derive the upper bound
\begin{align}
    p_1^{\text{kick}}\leq 2||P^{O_3}_1P^{O_1}_1||^2.
\end{align}
Unsurprisingly, the probability of finding a spin-up state depends directly on the probability that the particle could travel from $R_1$ to $R_3$ in time $t_3$.
\subsection*{The no-kick scenario}\label{sec:nokick}
We now start from the state
\begin{align}
    |\psi_{\text{no kick}}\rangle  &= |\psi _-\rangle \otimes |dd\rangle  \otimes |0\rangle_{O_2} \otimes|0\rangle _{O_3}
\end{align}
with
\begin{align}
    |\psi _-\rangle &= P_1^{O_1}\psi(x_1,x_2) - P_2^{O_2}\psi (x_2,x_1).
\end{align}
We obtain for the spin-up probability $p_1^{\text{no kick}}$ (see appendix \ref{app:tech})
\begin{align}
    p_1^{\text{no kick}} = &\frac{1}{2}\bigg[||P_{01}^{O_3}P_{11}^{O_2}(P_1^{O_1}\psi(x_1,x_2) - P_2^{O_1}\psi (x_2,x_1)) ||^2 \\
    &+ ||P_{10}^{O_3}P_{11}^{O_2}(P_1^{O_1}\psi(x_1,x_2) - P_2^{O_1}\psi (x_2,x_1)) ||^2  \bigg]
\end{align}
which, after some manipulation and applying the triangle inequality, turns in to an upper bound
\begin{align}
    p_1^\text{no kick} \leq \bigg[ ||(1-P^{O_3}_1)P^{O_2}_1P^{O_1}_1|| \cdot ||P^{O_3}_2 P^{O_2}_2|| +||P^{O_3}_2P^{O_2}_2P^{O_1}_2||\bigg]^2
\end{align}
Note that for the operator norms, it does not matter whether we use the first or second spatial Hilbert space for computing them; we have used differing indices to remain consistent with notation.
\section{Realizations}\label{sec:realization}
Let us now specialize to a particular unitary evolution in one space dimension. We will suppose that the particles are free, aside from the measurements and kicks occurring in the scenario. Thus the evolution is given by the Hamiltonian
\begin{align}
    H_p = -\frac{1}{2m}\frac{\partial^2}{\partial x^2}
\end{align}
for each particle. We use the simplest, nearest-neighbor discretization on an infinite evenly spaced grid to obtain 
\begin{align}
    H = -J\sum_n |n\rangle\langle n+1| + |n+1\rangle\langle n| -2|n\rangle\langle n|
\end{align}
where $J = \frac{1}{2ma^2}$, $a$ is the lattice spacing. The group velocity $v_{max} = 2Ja$, and we also write $z_i = v_{max}t_i/a$. The details are found in appendix \ref{app:tech}. The propagator between lattice points $n$ and $m$ is
\begin{align}
    \langle m | U(t) |n\rangle = i^{m-n}J_{m-n}(z),
\end{align}
where $J_n$ are Bessel functions of the first kind. Though this Hamiltonian is quite standard, the details for deriving it are in appendix \ref{app:tech}. We take regions 1 and 3 to be single points, call them $|n\rangle$ and $|m\rangle$ respectively, and the intermediate measurement projection happens on some subset $R\subset R_2$.

\begin{figure}
    \centering
    \includegraphics[width=0.95\linewidth]{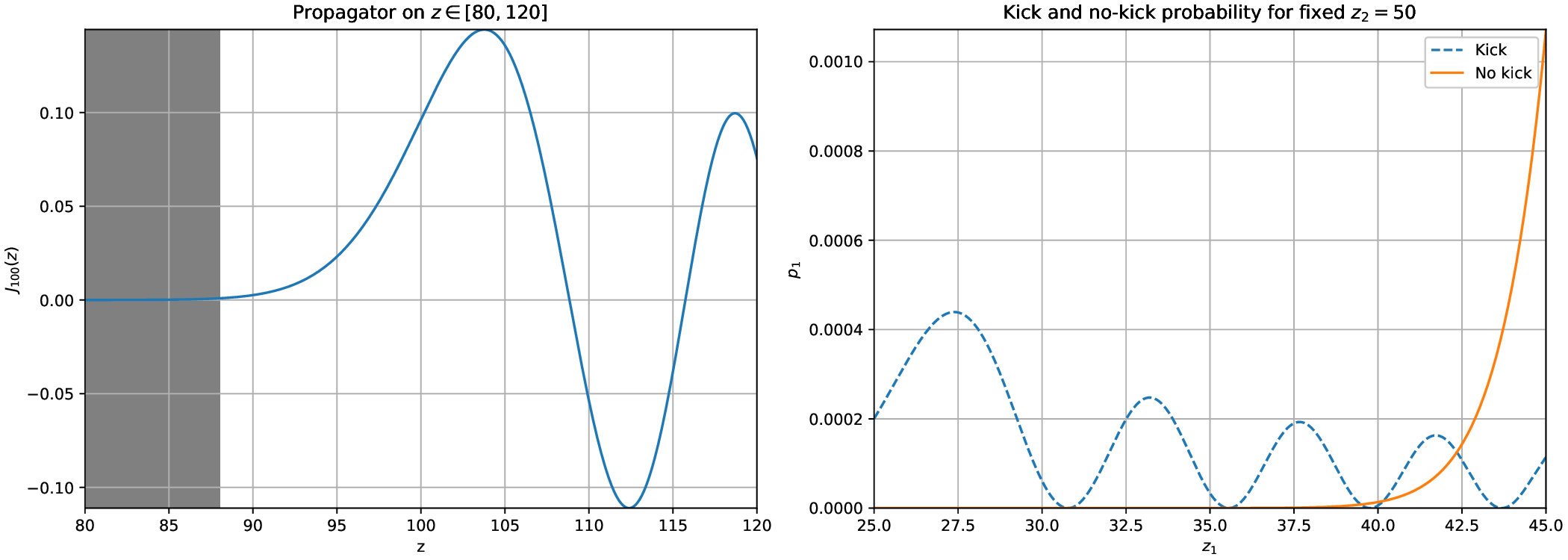}
    \caption{On the left, the behavior of the propagator for order $|n-m|=100$, which is the Bessel function $J_{100}$. The grey area is roughly the "no-signaling"\ region for this propagator. The amplitude is rapidly suppressed in the region $z<100$. This causes the theory to have suppressed signaling as the events become causally separated. On the right, the no-kick and kicked probabilities compared for a fixed choice of detector region and fixed $z_2$.}
    \label{fig:propagator}
\end{figure}

Note that in our model, two points $|m\rangle$ and $|n\rangle$ are said to be approximately causally separated under our time evolution when 
\begin{align}
    z < |m-n|
\end{align}
where $z$ is the argument of the propagator. To be sure that the probability really is negligible, we should choose $z$ considerably below $|m-n|$. This behavior evident when we look at the behavior of a Bessel function, as shown in figure \ref{fig:propagator}.

In practice, then, the constraints given by our model are as follows:
\begin{align}
    z_1 &> d(n,R), \label{eq:signaling1}\\
    z_2 &> d(m,R) ,\label{eq:signaling2}\\
    z_1+z_2 &< d(n,m)\label{eq:signaling3} 
\end{align}
where we take $d(n,R) = \inf \{ |n-m|\ |\ m\in R\}$. The first two constraints say that the measurement region is causally connected to $O_1$ and $O_3$; the last one says that $O_1$ and $O_3$ are not causally connected.

After a tedious calculation to be found in appendix \ref{app:tech}, we obtain
\begin{align}
    p_1^{\text{kick}} &\leq 2|J_{m-n}(z_1+z_2)|^2\label{eq:boundkick}\\
    p_1^{\text{no kick}} &\leq \bigg[ \sqrt{\sum_{k\in R} |J_{k-n}(z_1)|^2 - |\sum _{k\in R} J_{m-k}(z_2)J_{k-n}(z_1)|^2} \cdot \sqrt{\sum_{i\in R}|J_{m-i}(z_2)|^2}\nonumber \\
    &+ |\sum_{j\in R} J_{m-j}(z_2)J_{j-n}(z_1)|\bigg]^2 \label{eq:boundnokick}
\end{align}
The question is whether it is possible for bound \eqref{eq:boundnokick} to be lower than bound \eqref{eq:boundkick} for some measurement setup in $O_2$. Recall that we assumed from the very beginning that $\eqref{eq:boundkick}$ is so small as to be negligible. Formally, we might construct an equivalence class in which $p_1^{\text{kick}} \sim 0$. Therefore, if we can find \eqref{eq:boundnokick} smaller than \eqref{eq:boundkick}, there is no appreciable signaling in the model.

\begin{figure}
    \centering
    \includegraphics[width=0.9\linewidth]{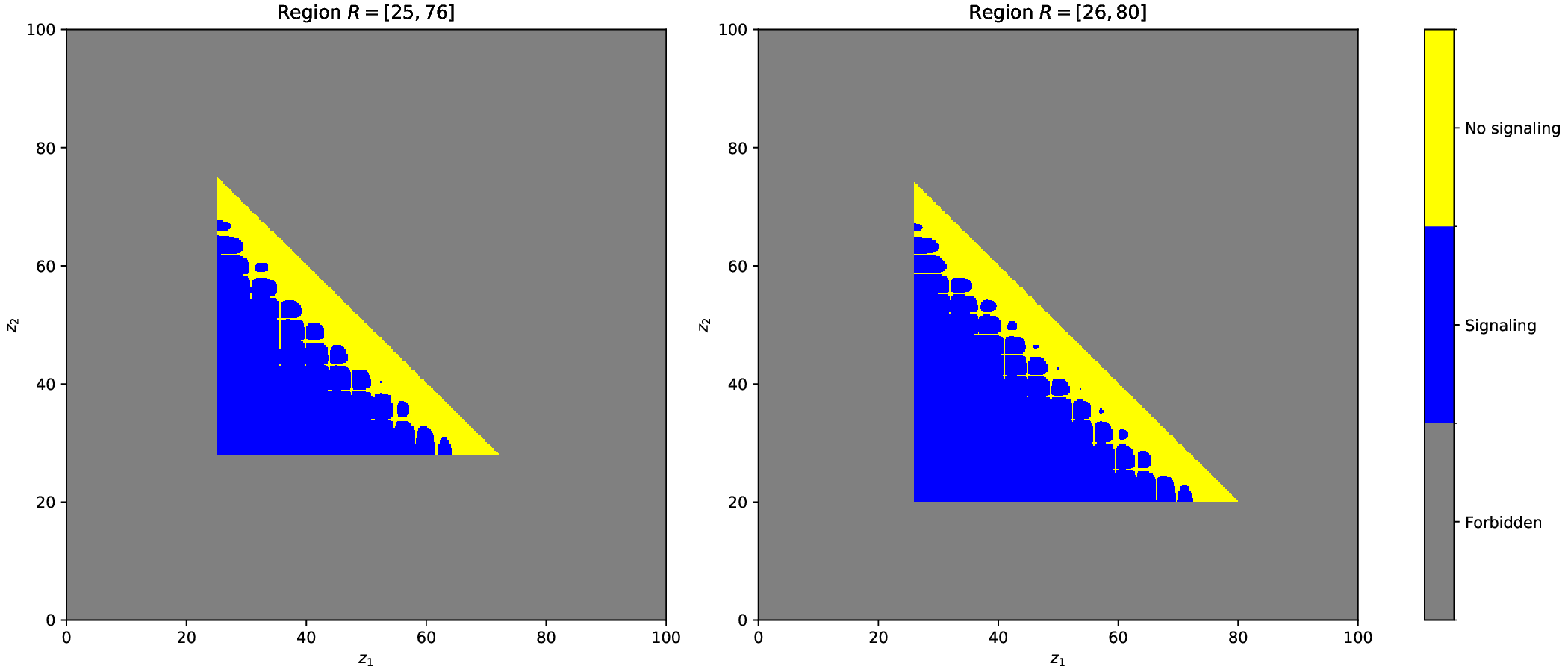}
    \caption{Signaling and no-signaling regions for two choices of measurement regions $R$ consisting of two gridpoints. Many other pairs of points -- and groups of three points -- can be found.}
    \label{fig:signalingheatmap}
\end{figure}
In fact, we can find such measurement setups; the no-signaling regions for certain choices of measurement setups are plotted in  If we take the measurement region to be continuous, then with the measurement scheme we have constructed it is either not possible or at any rate very difficult to find a region $R$ such that \eqref{eq:boundkick} is greater than \eqref{eq:boundnokick}. But if the measurement operator itself does not have to operate on the entire region $R_2$, but merely has to satisfy the signaling conditions \eqref{eq:signaling1}-\eqref{eq:signaling3}, then there are many possible choices of projectors that do this, consisting of disjoint projection regions.

Some intuition for this can be gained by observing that the Bessel functions have zeros are oscillatory and thus have many zeros. If one can find some set of lattice points $i$ such that, in combination with the choices of $z_1$ and $z_2$, the Bessel functions in \eqref{eq:boundnokick} are near their zeroes, then it is possible to obtain a very low bound indeed. Hence it is not automatic that a Sorkin-type scenario increases signaling, even in non-relativistic quantum mechanics.
\section{Discussion}\label{sec:discussion}
The foregoing analysis shows that the amount of signaling -- and indeed, its existence -- depends on the specific measurement scenario chosen, even for the non-relativistic theory. In this specific sense, it is thus heavily context-dependent. A practical consequence is that the use of continuous detectors -- rather than detectors consisting of isolated points -- tends to cause signaling. Physically, this might be taken to mean that unless the internal dynamics of the detector are modeled, we should not include internal points of the detector in our model.

We chose a rather simple measurement scheme, where the spatial part of the device consists of just a projection. It is not difficult to imagine more complicated measurement apparatuses. Each one of them would have to be analyzed on a case-by-case basis to determine what their effect is on signaling; sketching a few simple ones should convince the reader that, just as in our scenario, it is possible for it to both suppress and amplify the amount of signaling present.

Some points are worthy of further attention. We defined the no-signaling condition as obtaining a result below a certain minimum bound derived from the kicked scenario. In a non-relativistic theory, there is of course still some signaling. As we alluded to, one way to formalize this would be to create an equivalence class in which any number below the bound is equivalent to zero. Alternatively, we could construct a POVM that uses an "error rate"\ derived from this bound. Schematically, take the measurement of one spin with one spatial variable in region $R$. Then the error-free POVM for a spin measurement would be
\begin{align}
    E_0 &= (1-P^R)\otimes 1 + P^R\otimes |d\rangle\langle d|,\\
    E_1 &= P^R\otimes |u\rangle \langle u|. 
\end{align}
We can, however, introduce an error in the standard way:
\begin{align}
    E_0^{\text{err}} &= (1-P^R)\otimes 1 + P^R\otimes \sin^2\theta|d\rangle\langle d|,\\
    E_1^{\text{err}} &= P^R\otimes (|u\rangle \langle u|+\cos ^2\theta |d\rangle\langle d|) 
\end{align}
where $\theta$ is a parameter controlling the error. This would induce some chance of obtaining result $E_1$ even when the spin was actually down. In our model, the measurement is slightly more complicated, but the idea is the same: we could suppose our small amount of signaling merely represents noise in the detector, and set $\cos^2\theta$ equal to \eqref{eq:boundkick}. Using this POVM, we could check whether the kick and no-kick scenarios produce a distinguishable difference in signaling. However, this extra formalism would not significantly change our analysis.

What about the no-signaling conditions? Let us reproduce the no-signaling argument of Ghirardi, Rimini and Weber \cite{ghirardiGeneralArgumentSuperluminal1980}. In this argument, we consider an observable $\mathcal{L}_A$ on measurement apparatus $A$. Thus $\mathcal{L}_A$ only operates on the space of the apparatus $A$ (in our case, this would be the qubit in region $O_3$). In addition, there's another apparatus $B$ (for our purposes, kick in $O_1$) and two systems $U$ and $V$ which are the ones being probed by the apparatuses.

Calling $\hat{U}_{VB}$ the measurement interaction of $V$ with $B$, and $\hat{U}_{UA}$ the measurement interaction of $U$ with $A$, the authors postulate that the measurement $U_{UA}$ does not interfere with the system $V$ and vice versa. In other words, these operators commute. So the idea is that first, a measurement on $V$ is performed, and then on $U$, and finally the expectation value is computed. Under these assumptions, it is easy to see that the expectation value does not change according to whether the measurement on $V$ is performed or not, because the measurement operations commute.

In our case, it is clear that the conditions of the no-signaling argument are not fulfilled; the operation in $O_3$ never commutes with the one in $O_1$. One can quantify this by computing the norm between the projectors of $O_3$ and $O_1$:
\begin{align}
    ||[P^{O_3}_1,P^{O_1}_1]|| = |J_{n-m}(z_1+z_2)|\sqrt{1-J_{n-m}(z_1+z_2)^2}.
\end{align}
Hence, when considering the measurement operators we used, they "almost commute"\ if $O_1$ and $O_3$ are sufficiently distant. Assuming they commute will not cause a large difference in the end result, and that is what is reflected by our choice to ignore any signaling below the bound \eqref{eq:boundkick}.

When Sorkin's scenario is treated using a full quantum field theory, there is still a small amount of signaling present \cite{polo-gomezDetectorbasedMeasurementTheory2022}. In the QFT models the amount of signaling is much easier to quantify, as it is directly related to the volume of the detector itself. In the non-relativistic case the measurement scheme still makes a difference, but there is no simple pattern to it. This is easy to understand: QFT is a relativistic theory by construction, and will respect causality by default; only the addition of non-local operations can induce signaling.

We have made indirect contact with the algebraic approach used in many QFT measurement papers by expressing the results in terms of operator norms of projectors \cite{hellwigFormalDescriptionMeasurements1970,fewsterLecturesMeasurementQuantum2025}. We note that our approach generalizes easily to the bosonic case, as the bounds were derived primarily by using the triangle inequality, for which the sign of the state is irrelevant. Nevertheless, unlike the QFT analysis, ours depends heavily on the implementation of the measurement scheme.
\section{Conclusion}\label{sec:conclusion}
In conclusion, we have fully analyzed Sorkin's scenario in the non-relativistic case. We have expressed the amount of signaling in terms of operator norms of projectors and derived upper bounds on them explicitly. 

It turns out that the presence of signaling heavily depends on the implementation of the measurement scheme. Using these mathematical results, we have attempted to clarify the conceptual underpinnings of the scenario, similarly to the quantum field theory treatment found in \cite{polo-gomezDetectorbasedMeasurementTheory2022}. We have also provided several analytical results for the various bounds on probabilities and commutators.

In further work, it might be interesting to investigate a wide range of possible measurement implementations to see whether one could find an easily tunable one that never has signaling in a similar three-site scenario.
\section{Acknowledgements}
J.H. wishes to thank Vilho, Yrjö and Kalle Väisälä Foundation of the Finnish Academy of Science and Letters for financial support.
\bibliography{sn-bibliography}
\appendix

\section{Technical details} \label{app:tech}
\subsection{The spin-up probabilities}
\subsubsection{Kicked initial state}
We start from the state
\begin{align}
    |\psi_{\text{kick}}\rangle = [|\psi\rangle \otimes|ud\rangle 
    - |\tilde{\psi}\rangle\otimes|du\rangle]\otimes |0\rangle_{O_2}\otimes |0\rangle_{O_3},
\end{align}
where we define
\begin{align}
    |\psi\rangle := P_1^{O_1}\psi(x_1,x_2), \qquad
    |\tilde{\psi}\rangle := P_2^{O_1}\psi(x_2,x_1).
\end{align}
Thus our initial state at time $t=0$ is
\begin{align}
    |\psi _{\text{kick}}\rangle \langle \psi _{\text{kick}}| = \rho _0.
\end{align}

Let us say that the measurement on $O_2$ is performed at time $t=t_1$ and after a further $\Delta t = t_2$ has passed, the measurement on $O_3$ is performed. It is easy to see that the measurement in $O_2$ does nothing to the kicked state. Hence, let us trace out space $O_2$, which is trivial, and call the new state $\rho '_0 = |\psi '_{\text{kick}}\rangle\langle\psi '_{\text{kick}}| = \text{tr}_{O_2}[\rho_0]$. Thus, merely have to perform the measurement on $O_3$ at time $t_3 := t_1+t_2$. Define for convenience
\begin{align}
    |\sigma_1\rangle &= |ud\rangle\otimes|0\rangle_{O_3}, \\
    |\sigma_2\rangle &= |du\rangle\otimes|0\rangle_{O_3}.
\end{align}
and the operator $T_\alpha$
\begin{align}
    \begin{array}{c|cccc}
\alpha & 00 & 10 & 01 & 11\\\hline
T_\alpha|\sigma_1\rangle & |ud\rangle \otimes |0\rangle _{O_3}& |dd\rangle\otimes |1\rangle _{O_3} & |ud\rangle\otimes |0\rangle _{O_3} & |ud\rangle\otimes |0\rangle _{O_3}\\
T_\alpha|\sigma_2\rangle & |du\rangle\otimes |0\rangle _{O_3} & |du\rangle\otimes |0\rangle _{O_3} & |dd\rangle\otimes |1\rangle _{O_3} & |du\rangle\otimes |0\rangle _{O_3}
\end{array}
\end{align}
Then evidently
\begin{align}
    |\psi_\text{kick}^{O_3}\rangle=\hat{U}^{O_3}|\psi_{\text{kick}}'\rangle =  \sum_{\alpha \in \{00,01,10,11\}} P^{O_3}_\alpha | \psi\rangle \otimes T_\alpha |\sigma _1\rangle - P^{O_3}_\alpha|\tilde{\psi}\rangle \otimes T_\alpha |\sigma _2\rangle
\end{align}
and the density matrix
\begin{align}
    &\rho _{\text{kick}} = |\psi_{\text{kick}}^{O_3}\rangle\langle \psi_{\text{kick}}^{O_3}| \nonumber\\
    &= \sum _{\alpha,\beta \in \{00,10,01,11\}} \Big[
 P^{O_3}_\alpha |\psi\rangle\langle\psi| P^{O_3}_\beta
    \otimes T_\alpha|\sigma_1\rangle\langle\sigma_1|T_\beta^\dagger
 + P^{O_3}_\alpha |\tilde{\psi}\rangle\langle\tilde{\psi}| P^{O_3}_\beta
    \otimes T_\alpha|\sigma_2\rangle\langle\sigma_2|T_\beta^\dagger \nonumber\\
& - P^{O_3}_\alpha |\psi\rangle\langle\tilde{\psi}| P^{O_3}_\beta
    \otimes T_\alpha|\sigma_1\rangle\langle\sigma_2|T_\beta^\dagger- P^{O_3}_\alpha |\tilde{\psi}\rangle\langle\psi| P^{O_3}_\beta
    \otimes T_\alpha|\sigma_2\rangle\langle\sigma_1|T_\beta^\dagger
\Big] .
\end{align}

We can now immediately calculate from the form above that the probability for state $|1\rangle_{O_3}$ is
\begin{align}
    p_1^{\text{kick}} = || P^{O_3}_{10} |\psi\rangle||^2 +||P^{O_3}_{01}|\tilde{\psi}\rangle||^2 =  ||P^{O_3}_{10}P^{O_1}_1\psi(x_1,x_2)||^2 + ||P^{O_3}_{01}P^{O_1}_2\psi(x_2,x_1)||^2
\end{align}
since any term where the measurement device is in state 0 is eliminated, and where the norm is understood as the $l^2$ norm. The bound is easy to obtain by using the definition of operator norms:
\begin{align}
     ||P^{O_3}_{10}P^{O_1}_1\psi(x_1,x_2)||^2 + ||P^{O_3}_{01}P^{O_1}_2\psi(x_2,x_1)||^2 \leq ||P^{O_3}_{10}P^{O_1}_1||^2 + ||P^{O_3}_{01}P^{O_1}_2||^2.
\end{align}
Our choice of operator norm is the typical one, i.e. the supremum norm. In turn, we see that
\begin{align}
    ||P^{O_3}_{10}P^{O_1}_1||^2 &=  ||P^{O_3}_1P^{O_1}_1||^2 ||(1-P^{O_3}_2)||^2 = ||P^{O_3}_1P^{O_1}_1||^2, \\
    ||P^{O_3}_{01}P^{O_1}_2||^2 &= ||(1-P^{O_3}_1)P_2^{O_3}P_2^{O_1}||^2 = ||P^{O_3}_2P^{O_1}_2||^2.
\end{align}
By symmetry, these norms are the same, and thus we write
\begin{align}
    p_1^{\text{kick}} \leq 2||P^{O_3}_1P^{O_1}_1||^2.
 \end{align}
\subsubsection{No-kick initial state}
We now begin from the state
\begin{align}
    \rho_0^{\text{no kick}} &= |\psi _-\rangle \langle \psi _-|\otimes |dd\rangle \langle dd| \otimes |0\rangle_{O_2} \langle 0|_{O_2}\otimes|0\rangle _{O_3}\langle 0|_{O_3}\\
    |\psi _-\rangle &= P_1^{O_1}\psi(x_1,x_2) - P_2^{O_1}\psi (x_2,x_1)
\end{align}
After the measurement on $O_2$, we get
\begin{align}
    \rho_{O_2}^{\text{no kick}} = &(1-P^{O_2}_{11})|\psi_-\rangle\langle \psi_-| (1-P^{O_2}_{11})\otimes |dd\rangle \langle dd|\otimes |0\rangle_{O_2}\langle 0|_{O_2}\otimes |0\rangle_{O_3}\langle 0|_{O_3} \nonumber\\
    &+ P^{O_2}_{11}|\psi_-\rangle \langle \psi _-|P^{O_2}_{11}\otimes |\psi_{O_2}\rangle\langle \psi_{O_2}|\otimes |0\rangle_{O_3}\langle 0|_{O_3}
\end{align}
with $|\psi _{O_2}\rangle = \frac{1}{2}(|dd\rangle\otimes |0\rangle _{O_2} -|uu\rangle\otimes|0\rangle _{O_2} +|dd\rangle\otimes|1\rangle _{O_2} + |uu\rangle\otimes|1\rangle _{O_2})$.

For the measurement in $O_3$, we define
\begin{align}
    |\chi_0\rangle &= |dd\rangle\otimes|0\rangle_{O_2}\otimes|0\rangle_{O_3}, \\
    |\chi_1\rangle &= \frac{1}{2}\big(|dd\rangle\otimes|0\rangle_{O_2}\otimes|0\rangle_{O_3} - |uu\rangle \otimes|0\rangle_{O_2}\otimes|0\rangle_{O_3}\nonumber\\
    &+ |uu\rangle \otimes|1\rangle_{O_2}\otimes|0\rangle_{O_3}+ |dd\rangle\otimes|1\rangle_{O_2}\otimes|0\rangle_{O_3}\big),
\end{align}
and the transformed versions under $\hat M^{(i)}$:\newpage
\begin{align}
    |\chi_{1,10}\rangle &= \frac{1}{2}\big(|dd\rangle\otimes|0\rangle_{O_2}\otimes|0\rangle_{O_3} - |du\rangle\otimes|0\rangle_{O_2}\otimes|1\rangle_{O_3} \nonumber\\
    &+ |du\rangle\otimes|1\rangle_{O_2}\otimes|1\rangle_{O_3} + |dd\rangle\otimes|1\rangle_{O_2}\otimes|0\rangle_{O_3}\big), \\
    |\chi_{1,01}\rangle &= \frac{1}{2}\big(|dd\rangle\otimes|0\rangle_{O_2}\otimes|0\rangle_{O_3} - |ud\rangle\otimes|0\rangle_{O_2}\otimes|1\rangle_{O_3} \newline\\
    &+ |ud\rangle\otimes|1\rangle_{O_2}\otimes|1\rangle_{O_3} + |dd\rangle\otimes|1\rangle_{O_2}\otimes|0\rangle_{O_3}\big), \\
    |\chi_{1,00}\rangle &= |\chi_{1,11}\rangle = |\chi_1\rangle.
\end{align}

Then the final density matrix is
\begin{align}
    \rho_{O_3}^{\text{no kick}} =
    \sum_{\alpha,\beta \in \{00,10,01,11\}}
    \Big[
    P^{O_3}_\alpha A P^{O_3}_\beta \otimes |\chi_0\rangle\langle\chi_0|
    + P^{O_3}_\alpha B P^{O_3}_\beta \otimes |\chi_{1,\alpha}\rangle\langle\chi_{1,\beta}|
    \Big]
\end{align}
 with 
 \begin{align}
     A &= (1 - P^{O_2}_{11})|\psi_-\rangle\langle\psi_-|(1 - P^{O_2}_{11}), \\
    B &= P^{O_2}_{11}|\psi_-\rangle\langle\psi_-|P^{O_2}_{11}.
 \end{align}
We get by direct computation
\begin{align}
    p_1^{\text{no kick}} = &\frac{1}{2}\bigg[||P_{01}^{O_3}P_{11}^{O_2}(P_1^{O_1}\psi(x_1,x_2) - P_2^{O_1}\psi (x_2,x_1)) ||^2 \nonumber\\
    &+ ||P_{10}^{O_3}P_{11}^{O_2}(P_1^{O_1}\psi(x_1,x_2) - P_2^{O_1}\psi (x_2,x_1)) ||^2  \bigg].
\end{align}

We can now try to find a bound on this. By symmetry, the two parts are equal. Thus we can just bound one:
\begin{align}
    &||P_{01}^{O_3}P_{11}^{O_2}(P_1^{O_1}\psi(x_1,x_2) - P_2^{O_1}\psi (x_2,x_1)) ||\nonumber\\
    &\leq ||P^{O_3}_{01}P^{O_2}_{11}P^{O_1}_1\psi(x_1,x_2)|| + ||P^{O_3}_{01}P^{O_2}_{11}P^{O_1}_2\psi(x_2,x_1)||\nonumber\\
    &\leq ||P^{O_3}_{01}P^{O_2}_{11}P^{O_1}_1|| + ||P^{O_3}_{01}P^{O_2}_{11}P^{O_1}_2|| \nonumber\\
    &= ||(1-P^{O_3}_1)P^{O_2}_1P^{O_1}_1|| \cdot ||P^{O_3}_2 P^{O_2}_2|| + ||(1-P^{O_3}_1)P^{O_2}_1||\cdot ||P^{O_3}_2P^{O_2}_2P^{O_1}_2||.
\end{align}
We have by computation $||(1-P^{O_3}_1)P^{O_2}_1|| = 1$ if $O_2$ consists of two or more lattice sites. This is seen as follows. Take two projectors $P,Q$. Take
\begin{align}
    ||(1-P)Q|| = \sup_{||v||=1} ||(1-P)Qv||
\end{align}
with $v$ in $l^2$. This norm is exactly one if we can find $v$ such that $v\in\text{ran}(Q)$ and $Pv=0$. In other words, we must prove that $\text{ran}(Q)\cap \ker P \neq \{0 \}$. Take $S := \text{ran}(Q)$, and write $P = |\phi \rangle \langle \phi|$. Then for the following functional
\begin{align}
    f(x) = \langle \phi|x\rangle, \quad x \in S
\end{align}
we have by the rank nullity theorem
\begin{align}
    \dim (\ker (f)) = \dim (S) - \dim (\text{im}(f)) \geq \dim (S) -1 \geq 1
\end{align}
since we assumed the rank of $S\geq 2$ -- this is why we must assume there is more than one lattice site in the projector. So $\ker (f)$ contains a non-zero vector. But $\ker (f) = \{ x \in S | \langle \phi |x\rangle = 0\} = \text{ran}(Q) \cap \ker (P)$, and thus we have proved the claim.

Then 
\begin{align}
    &||P_{01}^{O_3}P_{11}^{O_2}(P_1^{O_1}\psi(x_1,x_2) - P_2^{O_1}\psi (x_2,x_1)) || \\
    &\leq ||(1-P^{O_3}_1)P^{O_2}_1P^{O_1}_1|| \cdot ||P^{O_3}_2 P^{O_2}_2|| +||P^{O_3}_2P^{O_2}_2P^{O_1}_2||.
\end{align}
Thus finally
\begin{align}
    p_1^\text{no kick} \leq \bigg[ ||(1-P^{O_3}_1)P^{O_2}_1P^{O_1}_1|| \cdot ||P^{O_3}_2 P^{O_2}_2|| +||P^{O_3}_2P^{O_2}_2P^{O_1}_2||\bigg]^2.
\end{align}
\subsection{The propagator}
The free Hamiltonian for a Schrödinger equation discretizes as
\begin{align}
    \hat{H} = -\frac{1}{2}\partial _x^2 \mapsto -J\sum_n |n\rangle\langle n+1| + |n+1\rangle\langle n| -2|n\rangle\langle n|
\end{align}
with $J = \frac{1}{2ma^2}$. Now define the discrete Fourier transform:
\begin{align}
    |k\rangle = \frac{1}{\sqrt{2\pi/a}}\sum _{n\in \mathbb{Z}} e^{ikna}|n\rangle.
\end{align}
By direct calculation
\begin{align}
    &\hat{H}|k\rangle \nonumber\\
    &= -J\sum_{n,m\in\mathbb{Z}} \bigg[|n\rangle\langle n+1| + |n+1\rangle\langle n| -2|n\rangle\langle n|\bigg]\frac{1}{\sqrt{2\pi/a}} e^{ikma}|m\rangle\nonumber\\
    &= -J \sum_{n}(e^{ik(n+1)a} + e^{ik(n-1)a}-2e^{ikna})\frac{|n\rangle}{\sqrt{2\pi/a}}\nonumber\\
    &= -J \sum_{n}e^{ikna}(e^{ika} + e^{-ika}-2)\frac{|n\rangle}{\sqrt{2\pi/a}}\nonumber\\
    &= -J\sum_n e^{ikna}(2\cos(ka)-2)\frac{|n\rangle}{\sqrt{2\pi/a}}\nonumber\\
    &= 2J(1-\cos(ka))|k\rangle .
\end{align}
Since $J = \frac{1}{2ma^2}$, we can expand as
\begin{align}
    2J(1-\cos(ka)) \approx \frac{(ka)^2}{2m} + \mathcal{O}((ka)^4).
\end{align}
From this, we see that our discrete Hamiltonian indeed approximates free Schrödinger evolution. However, we can ditch the constant diagonal term, because it merely leads to a global phase shift. Let us denote $\epsilon (k) = -2J\cos (ka)$

Now let's take two regions $R_1$ and $R_2$. Suppose those regions are just one lattice point for simplicity but without loss of generality. Then we want to know what probability we have for a particle initially localized in $R_1$ at $t=0$ to end up in $R_2$ at time $t$, i.e. we need to compute the propagator $\langle m|\hat{U}(t)|n\rangle$. The evolution operator is $U(t) = e^{-i\hat{H}_{tb}t}$ and the group velocity is $v_{max} = \text{max}_{k}\frac{dE}{dk}= 2Ja = 1/(ma)$ and we write $z = v_{max}t/a$.

Write the resolution of identity as
\begin{align}
    1 = \int _{-\pi /a}^{\pi/a}\frac{dk}{2\pi/a} |k\rangle \langle k|.
\end{align}
Shift the problem such that the ket is the vector of index 0 (with no loss of generality). Then:
\begin{align}
    \langle n|U(t)|0\rangle =\int _{-\pi /a}^{\pi/a}\frac{dk}{2\pi/a} \langle n|k\rangle\langle k|U(t)|0\rangle .
\end{align}
Now note that $\langle n|k\rangle = \frac{1}{\sqrt{2\pi /a}}e^{-ikna}$ and $\langle k |U(t)|0\rangle = e^{-i\epsilon (k)t}\langle k|0\rangle$. Combining these, we get
\begin{align}
    \langle n|U(t)|0\rangle = \int _{-\pi /a}^{\pi/a}\frac{dk}{2\pi/a} e^{ikna}e^{-i\epsilon (k) t}.
\end{align}
Now write $ka = \theta$, i.e. a change of variables. Then plugging in $\epsilon (k)$ we get
\begin{align}
    \langle n|U(t)|0\rangle = \frac{1}{2\pi }\int_{-\pi}^\pi d\theta e^{in\theta}e^{iz\cos (\theta)}.
\end{align}
Using the Jacobi-Anger expansion, we have 
\begin{align}
    e^{iz\cos (\theta)} = \sum_{m=-\infty}^\infty i^m J_m(z)e^{im\theta}.
\end{align}
For the propagator we obtain
\begin{align}
    \langle n|U(t)|0\rangle &= \frac{1}{2\pi} \sum_{m=-\infty}^\infty i^m J_m(z)\int _{-\pi}^\pi d\theta e^{i(n+m)\theta} = \sum _{m=-\infty}^\infty i^m J_m(z) \delta_{m,-n} \nonumber\\
    &= i^{-n}J_{-n}(z) = i^nJ_{n}(z).
\end{align}
Thus for general states
\begin{align}
    \langle n | U(t) | m\rangle = i^{n-m}J_{n-m}(z)
\end{align}
with $z = v_{max}t/a$.

This of course generalizes in the obvious way to N dimensions. Take $\vec{n} = (n_1,n_2,...n_d), \vec{m} = (m_1,m_2...m_d)\in \mathbb{Z}^d$ Then:
\begin{align}
    \langle \vec{n} | U(t) | \vec{m}\rangle = i^{\vec{n}-\vec{m}}J_{\vec{n}-\vec{m}}(z) = i^{\sum _i n_i-m_i}\prod _iJ_{n_i-m_i}(z).
\end{align}
\subsection{Operator norms}
We can use the propagator to compute operator norms. Let us calculate the most challenging term, as the others are computed in a similar way.
\begin{align}
    ||(1-P^{O_3})P^{O_2}P^{O_1}||^2 &= ||(1-P^{O_3})P^{O_2}|n\rangle\langle n||| | \nonumber\\
    &= |||\Phi '\rangle||\cdot ||\langle n|||\nonumber\\
    &= |||\Phi '\rangle||.
\end{align}
where $|\Phi '\rangle = (1-P^{O_3})P^{O_2}P^{O_1}|n\rangle$.
Thus we need to calculate
\begin{align}
    ||[1-U^\dagger(t_1+t_2) P^{R_3}U(t_1+t_2)]U^\dagger (t_1)P^{R_2}U(t_1)|n\rangle||.
\end{align}
With definitions
\begin{align}
    U^\dagger (t_1+t_2)|m\rangle &= |\varphi\rangle ,\\
    U(t_1)|n\rangle &= |\chi \rangle ,\\
     U^\dagger (t_1)\sum_{k\in R_2} |k\rangle\langle k|\chi\rangle &=|\phi\rangle ,
\end{align}
we find
\begin{align}
    ||[1-U^\dagger(t_1+t_2) P^{R_3}U(t_1+t_2)]U^\dagger (t_1)P^{R_2}U(t_1)|n\rangle||^2 &= || (1 - |\varphi \rangle \langle \varphi|)|\phi\rangle||^2 \nonumber\\
    &= ||\phi||^2 - |\langle \varphi | \phi \rangle |^2.
\end{align}
By computation, we find
\begin{align}
||\phi||^2 = \langle n| P^{R_2}(t_1) |n\rangle = \sum _{k\in R_2} |\langle k|\chi \rangle |^2 = \sum_{k\in R_2} |J_{k-n}(z_1)|^2
\end{align}
and for the overlap term
\begin{align}
    |\langle \varphi | \phi \rangle|^2 &= |\langle m| U(t_1+t_2) U^\dagger (t_1) \sum _{k\in R_2}|k\rangle \langle k| U(t_1)|n\rangle|^2 \nonumber\\
    &= |\sum _{k\in R_2} J_{m-k}(z_2)J_{k-n}(z_1)|^2
\end{align}
Hence we finally obtain
\begin{align}
    &||[1-U^\dagger(t_1+t_2) P^{R_3}U(t_1+t_2)]U^\dagger (t_1)P^{R_2}U(t_1)|n\rangle||^2 \nonumber\\
    &= \sum_{k\in R_2} |J_{k-n}(z_1)|^2 - |\sum _{k\in R_2} J_{m-k}(z_2)J_{k-n}(z_1)|^2.
\end{align}
Using similar techniques, we obtain:
\begin{align}
    ||P^{O_2}_1P^{O_1}_1|| &= \sqrt{\sum_{j\in R}|J_{j-n}(z_1)|^2},\\
    ||P^{O_3}_1P^{O_2}_1P^{O_1}_1|| &= |\sum_{j\in R} J_{m-j}(z_2)J_{j-n}(z_1)|,\\
    ||P^{O_3}_1P^{O_1}_1|| &= |J_{m-n}(z_1+z_2)|.
\end{align}
\end{document}